\documentclass[twocolumn,prb,aps,showpacs,superscriptaddress, amsmath, amssymb, notitlepage, longbibliography]{revtex4-2}
\usepackage[dvipdfmx]{graphicx} 
\usepackage{dcolumn}
\usepackage{bm}
\usepackage{braket}
\usepackage{amsfonts}
\usepackage{color}

\begin{document}
\preprint{APS/123-QED}
\title{Nonreciprocal heat transport in the Kitaev chiral spin liquid}
\author{Yoshiki Sano}
\email{sano@blade.mp.es.osaka-u.ac.jp}
\affiliation{Department of Materials Engineering Science, Osaka University, Toyonaka 560-8531, Japan}
\author{Daichi Takikawa}
\affiliation{Department of Materials Engineering Science, Osaka University, Toyonaka 560-8531, Japan}
\author{Masahiro O. Takahashi}
\affiliation{Department of Materials Engineering Science, Osaka University, Toyonaka 560-8531, Japan}
\author{\\Masahiko G. Yamada}
\affiliation{Department of Physics, School of Science, the University of Tokyo, Hongo, Bunkyo-ku 113-0033, Japan}
\author{Takeshi Mizushima}
\affiliation{Department of Materials Engineering Science, Osaka University, Toyonaka 560-8531, Japan}
\author{Satoshi Fujimoto}
\affiliation{Department of Materials Engineering Science, Osaka University, Toyonaka 560-8531, Japan}
\affiliation{Center for Quantum Information and Quantum Biology, Osaka University, Toyonaka 560-8531, Japan}
\date{\today}
\begin{abstract}
    Nonreciprocal transport, characterized by its direction-selective nature,
    holds significant potential for applications in various devices.
    In this study, we investigate nonreciprocal heat transport in Majorana systems,
    specifically focusing on the Kitaev chiral spin liquid under external magnetic fields.
    Our theoretical examination focuses on effects of open boundaries in which the Majorana edge modes exist,
    and the inversion symmetry is broken, which leads to the Dzyaloshinskii-Moriya interaction (DMI).
    Through perturbation theory, we demonstrate that DMI induces asymmetric hopping,
    resulting in the asymmetry of the Majorana band.
    The results of nonreciprocal heat currents are presented for various directions of external magnetic fields,
    and we discuss the relation between the current and the field-directions.
    The potential exists to manipulate both of the directions and magnitude of the nonreciprocal current
    by varying external magnetic fields and apply to heat transfer devices.
\end{abstract}
\maketitle
\section{Introduction}
    Nonreciprocal transport phenomena,
    characterized by their direction-selective transport,
    draw a lot of interest due to its significant potential for practical applications.
    One of the most known examples is the semiconductor p-n junction.
    The current for the direction $\xi$ is represented by
    \begin{align}
    	\mathcal{J}_{\xi}=\chi^{(1)}_{\xi}\mathcal{D}_{\xi}+\chi^{(2)}_{\xi}\mathcal{D}_{\xi}^{2}+\cdots,\label{eq:intro}
    \end{align}
    where the conductivity $\chi$ and the external drive field $\mathcal{D}$
    satisfy $\chi_{\xi}=\chi_{-\xi}$ and $\mathcal{D}_{\xi}=-\mathcal{D}_{-\xi}$.
    In particular, if the even-order conductivity is finite,
    the current magnitudes $|\mathcal{J}|$ differ between positive and negative directions.
    Nonreciprocal transport phenomena result from inversion and time-reversal symmetry breaking,
    which can cause higher-order effects that break Onsager reciprocal relations~\cite{Rikken2001,Tokura2018,Nagaosa2024}.
    Thus, the difference in the group velocities at $\bm{k}$ and $\bm{-k}$,
    a typical higher-order effect, can induce nonreciprocal transport phenomena.
    Nonreciprocal transport phenomena have been extensively studied
    in diverse materials,
    such as superconductors~\cite{Wakatsuki2017, Hoshino2018,  Itahashi2020, Ando2020},
    antiferromagnets~\cite{Hayami2022, Weizhao2022}, and topological insulators~\cite{Yasuda2019, Yasuda2020}.
    However, its exploration in Majorana systems remains
    relatively limited~\cite{Nakazawa2022}.
    
    Notably, the Kitaev honeycomb model stands as an exactly solvable Majorana system,
    which has attracted much attention because it acts as the quantum spin liquid in its ground state~\cite{Kitaev2006}.
    Jackeli and Khaliullin suggest that Mott insulators with the strong spin-orbit coupling and the $J_{\mathrm{eff}}=1/2$ moments
    have feasibility
    to realize the Kitaev spin liquid~\cite{Jackeli2009}.
    $4d$ and $5d$ transition metal compounds are proposed as candidates,
    for example, $\mathrm{Na_{2}IrO_{3}}$~\cite{Singh2010},
    $\mathrm{\alpha}$-$\mathrm{Li_{2}IrO_{3}}$~\cite{Singh2012}, $\mathrm{\alpha}$-$\mathrm{RuCl_{3}}$~\cite{Plumb2014},
    $\mathrm{H_{3}LiIr_{2}O_{6}}$~\cite{Geirhos2020} and so on.
    The $\mathrm{\alpha}$-$\mathrm{RuCl_{3}}$ is one of the promising candidates
    because the half-integer thermal Hall effect is observed,
    which is a signature of non-dissipative Majorana edge current~\cite{Kasahara2018,Yokoi2021}.

    \begin{figure}[b!]
        \includegraphics[width=\columnwidth]{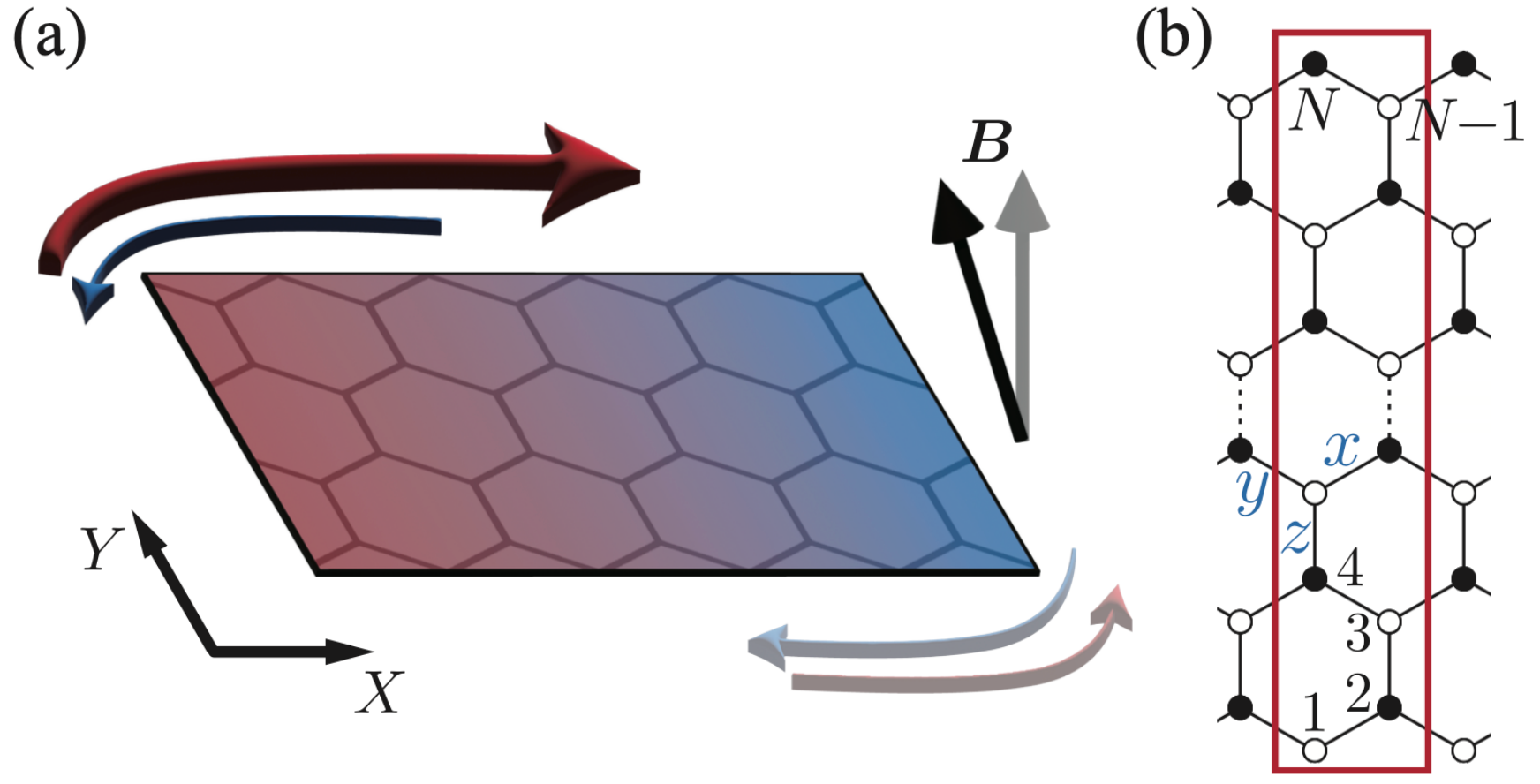}
        \caption{(a) The setup outline of the system.
        The temperature gradient, which is represented by the gradation of the red and blue colors,
        is applied for $X$ direction.
        (b) Unit cell of the system with open edges in $Y$ direction.
        We consider periodicity only for the $X$ direction.
        The empty and full circles represents even and odd sublattices.
        The flux sector is controllable by flipping the $\mathbb{Z}_{2}$ of each bonds.
        }
        \label{fig1}
    \end{figure}

    Several transport phenomena have been explored in the Kitaev chiral spin liquid.
    The itinerant Majorana fermions in the bulk act as carriers of spin~\cite{Minakawa2020, Nasu2020, Taguchi2021, Taguchi2022}
    and heat~\cite{Nasu2017, Pidatella2019}.
    Recently, the nonreciprocal heat transport in the Kitaev spin liquid
    by applying the staggard magnetic fields is discussed~\cite{Nakazawa2022}.
    Additionally, it has been suggested that the itinerant Majorana fermions
    at the edge play an important role in the spin current~\cite{Takikawa2022}.
    However, the nonreciprocal heat transport for the case with the open edge is still unclear,
    where the non-dissipative heat current exist and the inversion symmetry is broken.
    Also, the Dzyaloshinskii-Moriya interaction (DMI)~\cite{Dzyaloshinsky1958,Moriya1960},
    which is induced by the broken inversion symmetry,
    has a significant impact in the Kitaev honeycomb model~\cite{GangChen2019, Raiko2020, Jang2021, Furuya2024}.

    In this paper, we investigate the nonreciprocal heat current,
    which arises from chiral Majorana edge states, and can be manipulated by varying external magnetic fields.
    We conduct the calculation for the system which has the thermal gradient
    with consideration of the open edge [see Fig.~\ref{fig1}(a)].
    The unit cell is extended for the upper and lower sides
    to be the open boundaries, shown in Fig.~\ref{fig1}(b).
    We consider the DMI at edges due to the broken inversion symmetry at open boundaries.
    Thereby, asymmetric hoppings and asymmetric Majorana bands appear.
    The heat current is evaluated by using Boltzmann transport theory.
    As a result, we find that the nonreciprocal heat current depends on the direction of the magnetic fields.
    
    The organization of this paper is as follows.
    In Sec.~\ref{sec:Model}, we present the model and the basic formulation for 
    the calculation of transport properties focusing on effects of chiral Majorana edge states. 
    In Sec.~\ref{sec:Result}, we show the results of asymmetric band structures,
    heat currents and conductivities.
    We also discuss the relation between the nonreciprocal currents and the directions of magnetic fields.
    In Sec.~\ref{sec:Discussion}, we quantitatively estimate conductivities and thermal currents.
    The conclusion is given in Sec.~\ref{sec:Conclusion}.

\section{Model}\label{sec:Model}
    \subsection{Kitaev spin liquid with edge}\label{subsec:Kitaev}
        In this paper, we focus on the isotropic Kitaev model under a magnetic field, expressed as follows:
        \begin{align}
            \begin{split}
                H_{\mathrm{K}} + H^{(3)}_{\mathrm{h}}=-J\sum_{\gamma=x,y,z}\sum_{\langle jk\rangle_{\gamma}}S^{\gamma}_{j}&S^{\gamma}_{k}\\
                -\frac{3!h_{x}h_{y}h_{z}}{\Delta^{2}}\sum_{\langle\!\langle jkl\rangle\!\rangle}&S^{x}_{j}S^{y}_{k}S^{z}_{l},\label{eq:Kitaev}
            \end{split}
        \end{align}
        where $S^{\gamma}_{j}$ is the $\gamma~(=x,y,z)$ component of an $S=1/2$ spin operator at site $j$.
        To avoid confusion, we adopt $x$, $y$ and $z$ as the orientations of the spin space,
        while $X$ and $Y$ denote the orientations of the real space, as shown in Fig.~\ref{fig1}.
        The initial term in Eq.~\eqref{eq:Kitaev} represents the Kitaev interaction,
        characterized as an Ising-type interaction only for the nearest-neighbor (NN) sites.
        The notation $\langle jk \rangle_{\gamma}$
        indicates the summation over NN sites connected by $\gamma$-bonds.
        Here, we set the interaction strength $J$ as the unit of energy
        with $J>0$, but our results are independent of the sign of $J$.
        The second term in Eq.~\eqref{eq:Kitaev} is a perturbation term induced
        by the Zeeman coupling $H_{\mathrm{h}} = -\sum_{j}h_{\gamma}S^{\gamma}_{j}$.
        The mean flux excitation energy is $\Delta$, and we set $\Delta=0.065J$.
        The coefficient $3!$ is due to the permutation of three spin operators arising from the perturbative processes.
        The notation $\langle\!\langle jkl \rangle\!\rangle$ denotes the summation exclusively 
        over triplets of sites $j,k,$ and $l$ that are aligned and interconnected.
        Eq.~\eqref{eq:Kitaev} can be exactly solved using the well-known Majorana representation,
        $S^{\gamma}_{j}=\frac{i}{2}b^{\gamma}_{j}c_{j}$.
        Here, both $b_j^\gamma$ and $c_j$ satisfy the Majorana condition $\eta^\dagger = \eta$ and
        $\{\eta_j^\alpha,\eta_k^\beta\}=\delta_{jk}\delta^{\alpha\beta}$ (with $\eta_j^\alpha=b_j^x,b_j^y,b_j^z,c$).
        We start by considering a thermal current in the ground state,
        i.e.,~$u_{j,k}=ib^{\gamma}_{j}b^{\gamma}_{k}=1$~($j\in{\textrm{even sublattice}}$,
        $k\in{\textrm{odd sublattice}}$) for all bonds.
        Each of the first and second terms in Eq.~\eqref{eq:Kitaev}
        transforms into
        a NN hopping and a next nearest-neighbor (NNN) hopping, respectively.
        When $h_{x}h_{y}h_{z} \neq 0$, the Chern number
        takes values $\nu=\pm1$, reflecting the sign of $h_xh_yh_z$,
        and the non-dissipative chiral current is realized at the edge.
        
        In order to consider the zig-zag edge, we adopt the unit cell depicted in Fig.~\ref{fig1}(b).
        Here, we introduce additional Majorana sites $c_{0}$ and $c_{N+1}$,
        each of which is located on one side of the edge, due to the presence of an unpaired $b$-Majorana fermion $b^z$.
        The additional Majorana fermions on both edges can couple with the nearest $c$-Majorana fermion on the same site
        through the magnetic field as follows:
        \begin{align}
            \begin{split}
                H_{\mathrm{edge}} &= -h_{z}\sum_{l}(S^{z}_{l,1} + S^{z}_{l,N})\\
                &=-\frac{h_{z}}{2}\sum_{l}(ic_{l,0}c_{l,1}+ic_{l,N+1}c_{l,N}),
            \end{split}
        \end{align}
        where $l$ and $m$ in $c_{l,m}$ are the indices of unit cells and sites, respectively.
        \begin{figure}[t!]
            \includegraphics[width=\columnwidth]{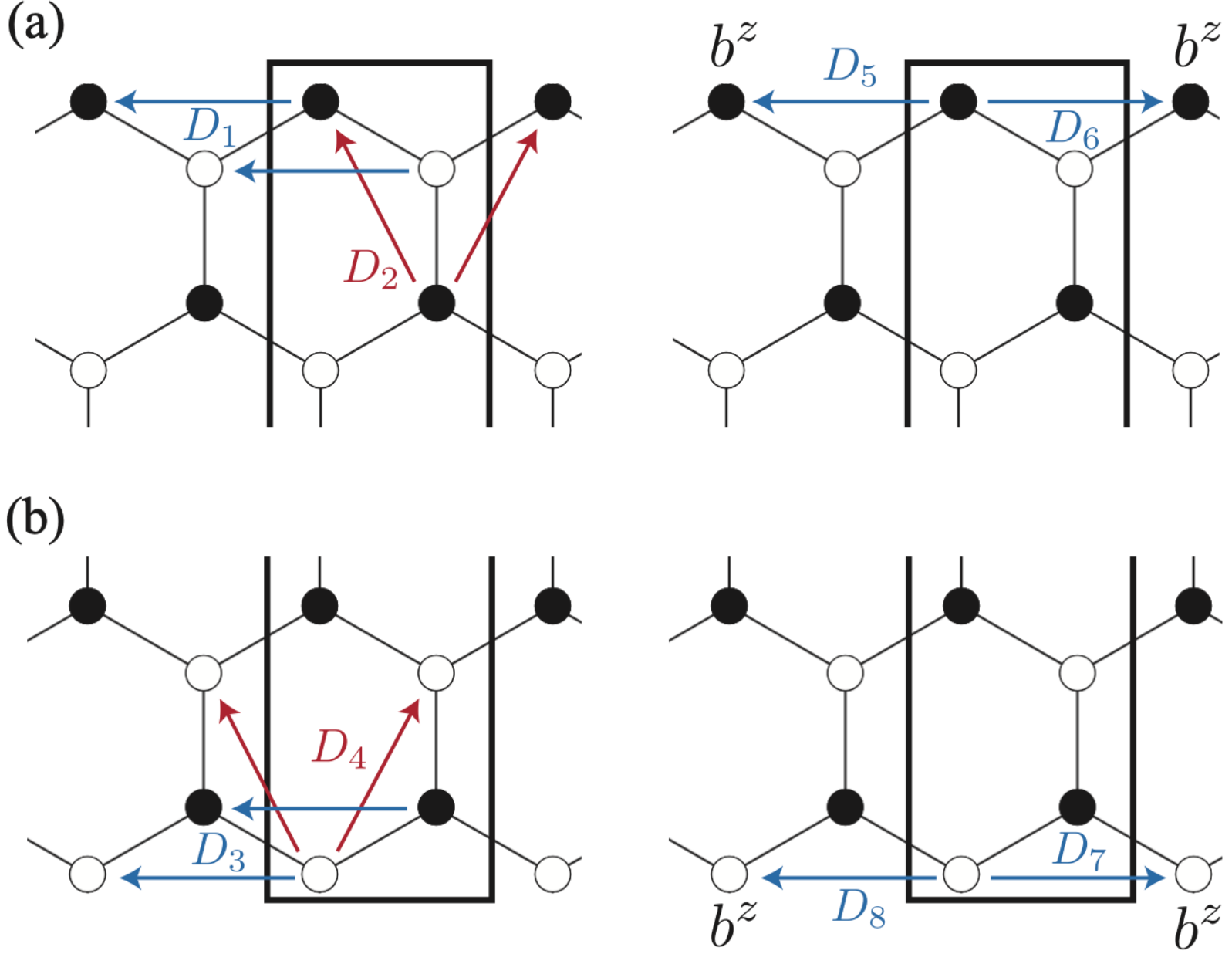}
            \caption{(a) DMI hopping directions of the upper edge,
            the indices of the unit cell and the site are
            $D_{1}:(l, N)\rightarrow(l-1, N)$ and $(l, N-1)\rightarrow(l-1, N-1)$,
            $D_{2}:(l, N-2)\rightarrow(l, N)$ and $(l, N-1)\rightarrow(l-1, N-1)$,
            $D_{5}:(l, N)\rightarrow (l-1, N+1)$, $D_{6}:(l, N)\rightarrow (l+1, N+1)$.
            (b) DMI hopping directions of the lower edge, the indices of the unit cell and the site are
            $D_{3}:(l, 1)\rightarrow (l-1, 1)$ and $(l, 2)\rightarrow(l-1, 2)$,
            $D_{4}:(l, 1)\rightarrow(l,3)$ and $(l, 1)\rightarrow(l-1,3)$,
            $D_{7}:(l, 1)\rightarrow (l+1, 0)$, $D_{8}:(l, 1)\rightarrow (l-1, 0)$.
            $D_{5}$, $D_{6}$, $D_{7}$ and $D_{8}$ are hopping
            from the $c$ operator to the edge $b$ operator. The black square represents the unit cell.
            }
            \label{fig2}
        \end{figure}
        Moreover, the DMI at the edges is taken into account
        due to the broken inversion symmetry at these boundaries.
        The DMI is expressed as,
        \begin{align}
            H_{\mathrm{DM}}=-\sum_{\langle jk\rangle'}\bm{D} \cdot [\bm{S}_{j} \times \bm{S}_{k}], \label{eq:DMspin}
        \end{align}
        where the summation is restricted to NN sites ($j\in{\textrm{even sublattice}}$, $k\in{\textrm{odd sublattice}}$),
        with at least one site belonging to the edge.
        For simplicity, we assume that the inversion symmetry in the bulk is not broken,
        considering the DMI only at the edges.
        In the second-order perturbation combining the DMI and magnetic fields, the effective Majorana Hamiltonian is given as follows:
        \begin{align}
            H^{(1)}_{\mathrm{DM}} = -\frac{1}{4\Delta}\sum_{p=1}^{8}\sum_{(l,m)(l',n)_{p}}D_{p}ic_{l,m}c_{l',n}, \label{eq:effDM}
        \end{align}
        where the strength of the couplings are given as,
        \begin{align}
            \begin{split}
                D_{1} &= D_{3} = h_{x}D_{x}-h_{y}D_{y},\\
                D_{2} &= D_{4} = h_{z}D_{z},\\
                D_{5} &= -D_{7} = 2h_{x}D_{y},\\
                D_{6} &= -D_{8} = -2h_{y}D_{x}, \label{eq:DMhop}
            \end{split}
        \end{align}
        The directions of hoppings are shown in Fig.~\ref{fig2} (detailed derivations are provided in Appendix~\ref{sec:AA}).
        Note that the interactions in Eq.~\eqref{eq:effDM} are NNN interactions.
        In this study, we set $\bm{D}=(0.1, 0.1, 0.1)$.

        \begin{figure*}[t!]
            \includegraphics[width=180mm]{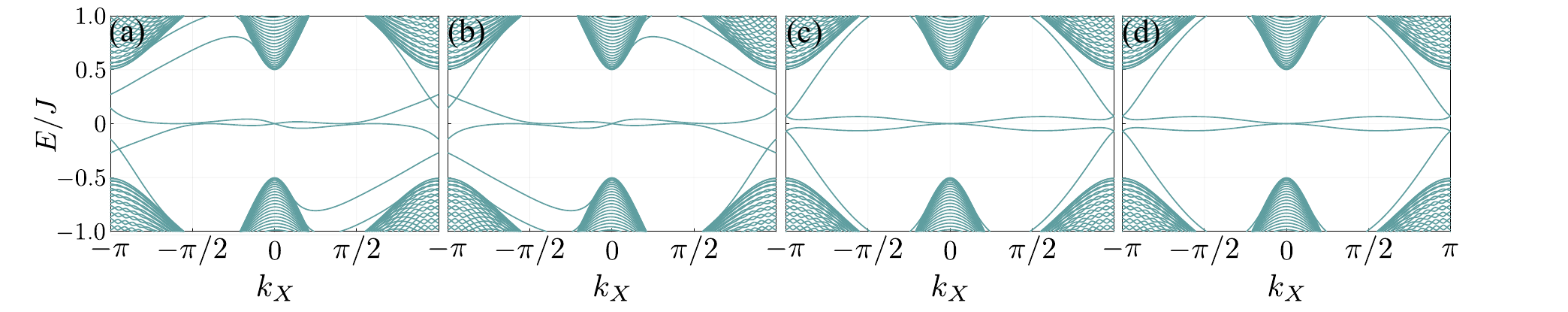}
            \caption{Distorted band structures in (a) $\bm{h}=(0.1, -0.1, 0.1)$, (b) $\bm{h}=(-0.1, 0.1, 0.1)$,
            (c) $\bm{h}=(0.1, 0.1, 0.1)$, (d) $\bm{h}=(0.1, 0.1, -0.1)$.
            (a) and (b) are asymmetric and inverted. (c) and (d) are numerically same.
            }
            \label{fig3}
        \end{figure*}

    \subsection{Total effective Hamiltonian}\label{subsec:Total}
        Finally, the total effective Hamiltonian discussed in this paper is summarized as follows:
        \begin{align}
            \begin{split}
                H_{\mathrm{tot}} &= H_{\mathrm{K}} + H^{(3)}_{\mathrm{h}} + H_{\mathrm{edge}} + H^{(1)}_{\mathrm{DM}}\\ 
                &= \frac{1}{4}\sum_{l,l',m,n}c_{l,m}A_{(l,m)(l',n)}c_{l',n}.\label{eq:tot1}
            \end{split}
        \end{align}
        Through the Fourier transformation of Majorana operators defined by
        \begin{align}
            c_{l,m} = \sqrt{\frac{2}{N_{\mathrm{unit}}}}\sum_{k_{X}}e^{ilk_{X}}c_{k_{X}, m},
        \end{align}
        where $N_{\mathrm{unit}}$ is the numbers of unit cells in a row along $X$ direction.
        Eq.~\eqref{eq:tot1} is converted into 
        \begin{align}
            H_{\mathrm{tot}} = \frac{1}{4}\sum_{k_{X},m,n}c_{-k_{X},m}\tilde{A}_{m,n}(k_{X})c_{k_{X},n},
        \end{align}
        where
        \begin{align}
            \tilde{A}_{m,n}(k_{X}) := 4\sum_{l,l'}e^{-ik_{X}(l-l')}A_{(l,m)(l',n)}.\label{eq:matrix}
        \end{align}
        Here the basis of $\tilde{A}$ is $N+2$ -dimensional $(c_{0}, c_{1}, \cdots, c_{N}, c_{N+1})$.

    \subsection{Boltzmann transport equation}\label{subsec:Boltzmann}
        The heat current and heat conductivities are calculated through the utilization of the Boltzmann transport equation.
        The heat current $\mathcal{J}$ and the heat conductivity $\kappa$ are described up to the second order as,
        \begin{align}
            \mathcal{J} = \kappa^{(1)}\left(-\frac{\partial T}{\partial X}\right)+\kappa^{(2)}\left(\frac{\partial T}{\partial X}\right)^{2}+ \cdots,\label{eq:current}
        \end{align}
        where
        \begin{align}
            \begin{split}
                \kappa^{(1)}&=\frac{2\tau}{\Omega}\sum_{k_{X}}\sum_{E_{n}<0}\frac{\partial f(E_{k_{X}n})}{\partial T}E_{k_{X}n}v_{k_{X}n}^{2},\\
                \kappa^{(2)}&=\frac{2\tau^{2}}{\Omega}\sum_{k_{X}}\sum_{E_{n}<0}\frac{\partial^{2} f(E_{k_{X}n})}{\partial T^{2}}E_{k_{X}n}v_{k_{X}n}^{3}.\label{eq:kappa}    
            \end{split}
        \end{align}
        $v_{k_{X}n}$ represents the group velocity in the $X$ direction
        ($v_{k_{X}n}=\partial E_{k_{X}n}/{\partial k_{X}}$) and $n$ is the index of Majorana bands.
        The contributions from both the Majorana edge bands and the bulk
        are summed up in the second summation of Eq.~\eqref{eq:kappa}.
        $\Omega$ denotes the system size,
        $\Omega=N k_{\mathrm{mesh}}$, where $k_{\mathrm{mesh}}=1/N_{\mathrm{unit}}$. (A unit of length is a lattice constant.)
        It is important to note that the energy current is equivalent to the heat current,
        since the chemical potential of Majorana particles can be set to zero~\cite{Nasu2017}.
        We adopt the Boltzmann constant value $k_{\mathrm{B}}=1$,
        and $f$ represents Fermi distribution function, $f(E_{k_{X} n}) = (\exp(E_{k_{X}n}/T)+1)^{-1}$.
        Here, $\tau$ is the relaxation time.       
        Assuming that backscattering of Majorana edge modes occurs
        mediated by the bulk excitations at finite temperature,
        we employ the relaxation time approximation and set $J\tau=1$.
        The second-order conductivity is nonlinear conductivity,
        which provides the nonreciprocal contribution.

\section{Results}\label{sec:Result}

    \subsection{Band structure}
    \begin{figure}[b!]
        \includegraphics[width=\columnwidth]{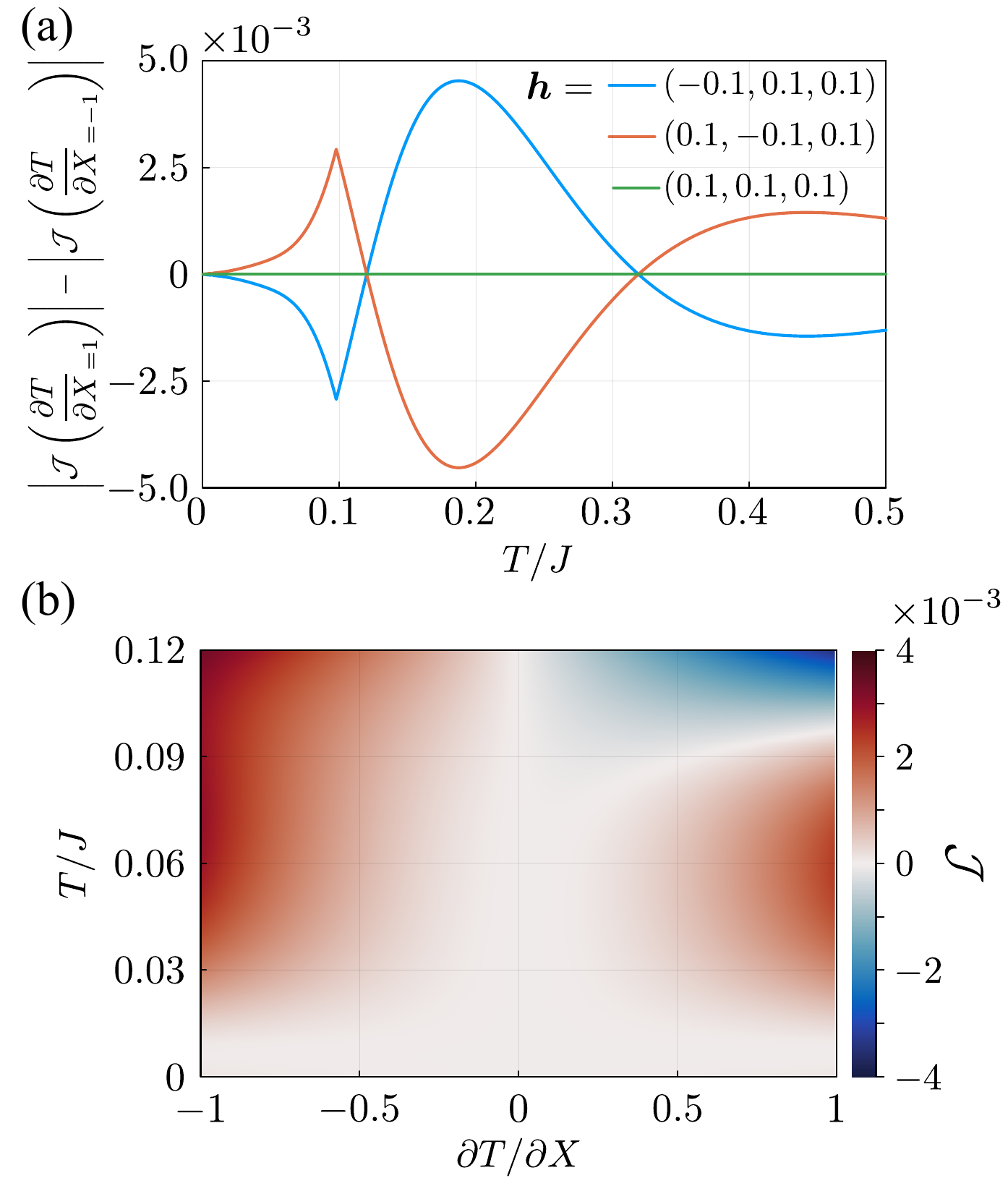}
        \caption{(a) The nonreciprocity of the current,
        which is represented by the difference in heat currents for the oppositely applied thermal gradients,
        $|\mathcal{J}(\partial T/ \partial X =1)|-|\mathcal{J}(\partial T/ \partial X =-1)|$.
        The blue, red and green lines correspond to
        $\bm{h}=(-0.1, 0.1, 0.1)$, $\bm{h}=(0.1, -0.1, 0.1)$ and $\bm{h}=(0.1, 0.1, 0.1)$, respectively.
        Changing magnetic direction inverts the direction of the current, corresponding to the band inversion.
        (b) The color map of the nonreciprocal heat current focused on
        the low temperature range for $\bm{h}=(-0.1, 0.1, 0.1)$.
        At $T/J < 0.10$, strong nonreciprocal current occurs.}
        \label{fig4}            
    \end{figure}
    We, first, show the band structures obtained by diagonalizing Eq.~\eqref{eq:matrix} in Figs.~\ref{fig3}(a)-(d).
    In Figs.~\ref{fig3}(a)~and~\ref{fig3}(b), the band structures are asymmetric,
    while, in Figs \ref{fig3}(c)~and~\ref{fig3}(d), the band structures are distorted by the DMI but still symmetric.
    The degree of the asymmetry depends on the edge-parallel hoppings
    ($D_{1}$, $D_{3}$, $D_{5}$, $D_{6}$, $D_{7}$ and $D_{8}$),
    while the other terms ($D_{2}$ and $D_{4}$), i.e., $h_{z}$ do not affect it.
    Additionally, the band structures shown in Figs.~\ref{fig3}(a)~and~\ref{fig3}(b)
    are related to each other via  
    the inversion operation ($k_{X}\rightarrow -k_{X}$)
    because the sign changes of magnetic fields, ``$h_{x} \rightarrow -h_{x}$ and $h_{y}\rightarrow -h_{y}$",
    results in the change in the signs of the edge-parallel hoppings.
    The band structures shown in Figs.~\ref{fig3}(c)~and~\ref{fig3}(d) are the same,
    when the hoppings $D_{1}$ and $D_{3}$ are zero under the condition $h_{x} = h_{y}$.
    Notably, it is advantageous for $h_{x}$ and $h_{y}$ to have opposite signs to maximize edge-parallel hoppings
    and manifest the effects of the edges.

    \begin{figure}[b!]
        \includegraphics[width=\columnwidth]{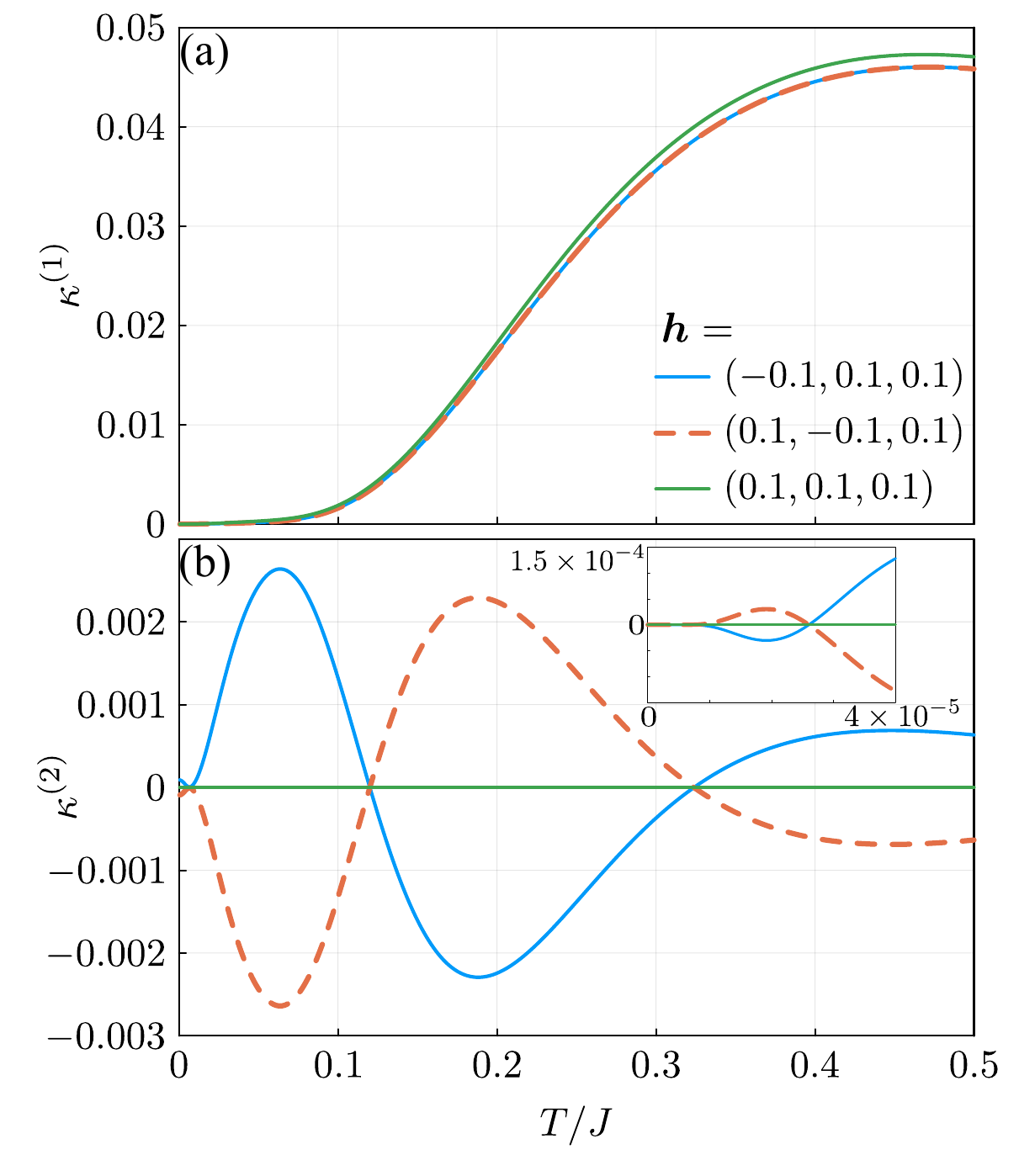}
        \caption{(a) Results of the calculation of $\kappa^{(1)}$ in Eq.~\eqref{eq:kappa}.
        The red broken line overlaps with the blue line.
        (b) Results of the calculation of $\kappa^{(2)}$ in Eq.~\eqref{eq:kappa}.
        Each of blue line, red broken line and green line represents
        $\bm{h}=(-0.1, 0.1, 0.1)$, $\bm{h}=(0.1, -0.1, 0.1)$ and $\bm{h}=(0.1, 0.1, 0.1)$.
        For $\kappa^{(2)}$, the sign is inverted between $\bm{h}=(-0.1, 0.1, 0.1)$ and $\bm{h}=(0.1, -0.1, 0.1)$.
        In the low temperature ($T/J \lesssim 0.09$), $\kappa^{(2)}$ is larger
        than $\kappa^{(1)}$ in $\bm{h}=(-0.1, 0.1, 0.1)$ and $\bm{h}=(0.1, -0.1, 0.1)$.
        The inset shows the same data at the extreme low temperature.
        The value oscillates with the respect to the temperature and reaches zero.}
        \label{fig5}            
    \end{figure}

    \subsection{Nonreciprocal current and conductivity}
    In Fig.~\ref{fig4}(a), we show the nonreciprocity of the current,
    defined as the difference in heat currents for the oppositely applied thermal gradients,
    $|\mathcal{J}(\partial T/ \partial X =1)|-|\mathcal{J}(\partial T/ \partial X =-1)|$.
    Each of blue, red and green line represents
    $\bm{h}=(-0.1, 0.1, 0.1)$, $\bm{h}=(0.1, -0.1, 0.1)$ and $\bm{h}=(0.1, 0.1, 0.1)$.
    The blue and red lines exhibit the same magnitude with opposite signs,
    reflecting the asymmetry and the inversion of the band structures shown in Figs.~\ref{fig3}(a) and \ref{fig3}(b).
    Additionally, the nonreciprocity has two peaks at $T/J\simeq0.1$ and $0.2$.
    In contrast, the green line shows no nonzero value, corresponding to the symmetric band structure in Fig.~\ref{fig3}(c).
    In Fig.~\ref{fig4}(b), we show the color map of the heat current
    plotted as functions of temperature and temperature gradient.
    Note again that Eq.~\eqref{eq:current}
    includes the contributions from both upper and lower edges of the system,
    in contrast to previous studies, for example Ref.~\cite{Yasuda2020}.
    Thus, the nonreciprocal behaviors are not due to chiral character of the edge states,
    but arise from the asymmetric Majorana bands caused by the DMI.
    The magnetic fields $\bm{h}$ in Fig.~\ref{fig4}(b)
    correspond to those in Fig.~\ref{fig3}(b) and the blue line in Fig.~\ref{fig4}(a).
    The nonreciprocity reaches its peak at $T/J\simeq0.1$ in Fig.~\ref{fig4}(a),
    where the value of $|\mathcal{J}(\partial T/ \partial X=1)|$ turns negative.
    The pronounced nonreciprocal heat transport is observed at low temperature ($T/J < 0.10$),
    where the heat current flows in the one direction regardless of the heat gradient.
    It is important to note that other contributions to the heat current,
    such as phonon contributions, should be considered to quantitatively evaluate the nonreciprocity.
    However, we anticipate that the nonreciprocal heat transport
    mediated by Majorana edge modes can be clearly observed in the low temperature regime,
    despite considering contributions from both upper and lower edges.
    Indeed, even though $\kappa^{(2)}$ retains a finite value above $T\gtrsim0.10$,
    the reciprocal part $\kappa^{(1)}$ increases to offset the nonreciprocal contribution.
    The direction of the current differs between $\bm{h}=(0.1, -0.1, 0.1)$ and $\bm{h}=(-0.1, 0.1, 0.1)$ (same as Fig.~\ref{fig4}(b)),
    corresponding to the red and blue line in Fig.~\ref{fig4}(a).

    \begin{figure}[t!]
        \includegraphics[width=\columnwidth]{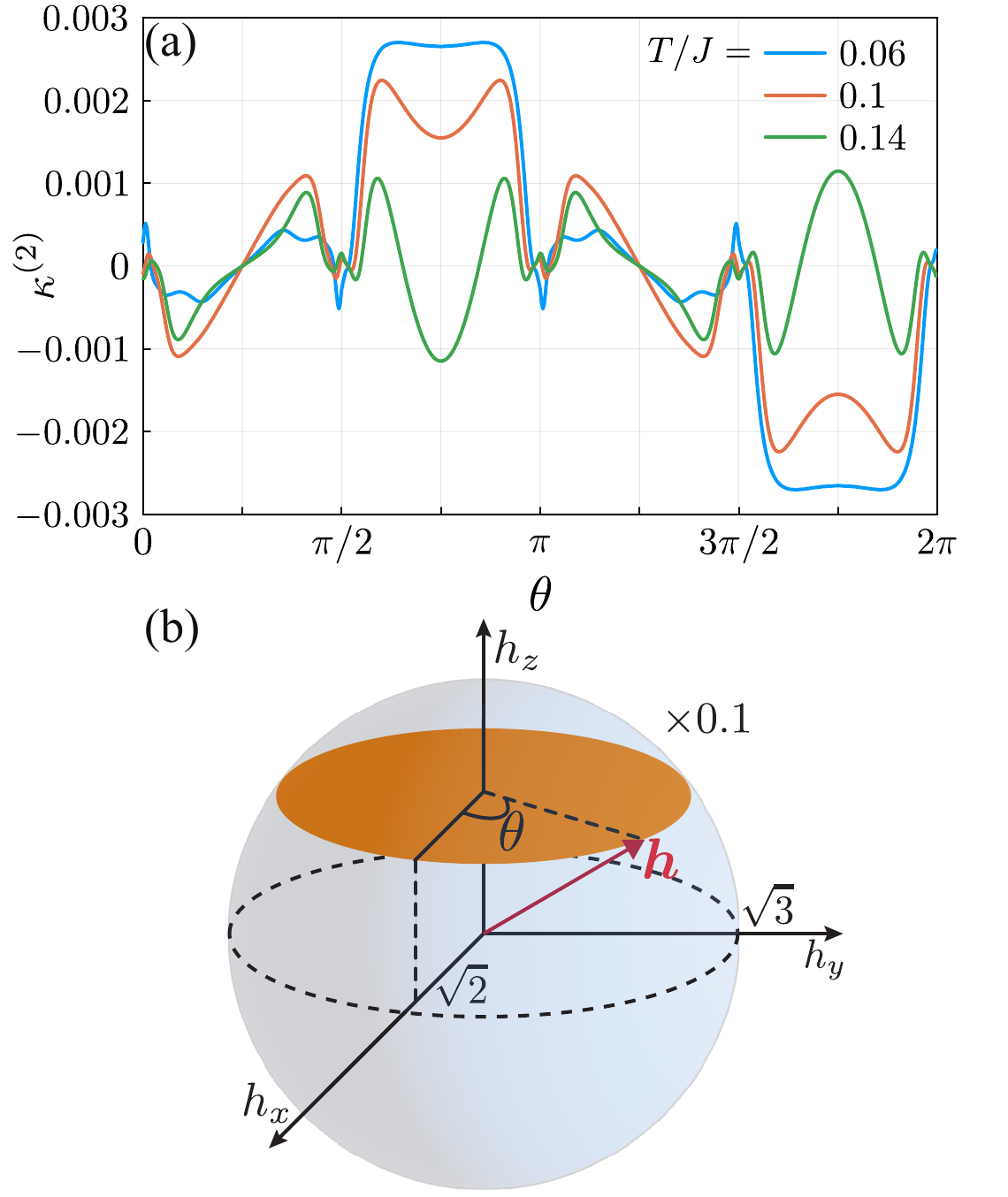}
        \caption{(a) The direction of magnetic field dependence of $\kappa^{(2)}$.
        The nonlinear conductivity is maximized at $\theta=3\pi/4, 7\pi/4$
        and vanishes at $\theta=\pi/4, 5\pi/4$.
        Each of blue, red and green line represents $T/J=0.06$, $T/J=0.1$ and $T/J=0.14$.
        (b) What the horizontal axis $\theta$ in (a) represents.
        We vary the magnetic field to keep $|\bm{h}|$ uniform.
        }
        \label{fig6}            
    \end{figure}

    We next explore the potential
    to manipulate the direction and magnitude of nonreciprocal heat current by varying external magnetic fields.
    For a detailed analysis, the results of the heat conductivities,
    $\kappa^{(1)}$ and $\kappa^{(2)}$ given by Eq.~\eqref{eq:kappa}, are shown in Figs.~\ref{fig5}(a) and \ref{fig5}(b).
    Each blue line, red broken line and green line corresponds to $\bm{h}=(-0.1, 0.1, 0.1)$,
    $\bm{h}=(0.1, -0.1, 0.1)$ and $\bm{h}=(0.1, 0.1, 0.1)$, respectively.
    In Fig.~\ref{fig5}(a), blue line and red broken line of $\kappa^{(1)}$ are numerically identical.
    However, for $\kappa^{(2)}$ in Fig.~\ref{fig5}(b), they have the same absolute value but the opposite sign
    due to the combined influence of the band distortion and orientation, governed by $v^{3}$ in Eq.~\eqref{eq:kappa}.
    The majority of contributions to the nonlinear conductivity come from the Majorana edge bands,
    as inversion symmetry is broken only at both edges in our model.
    In fact, $\kappa^{(2)}$ for $\bm{h}=(0.1, 0.1, 0.1)$ (shown as a green line in Fig.~\ref{fig5}(b))
    is numerically zero because the band structure is symmetric.
    
    Additionally, we examine the dependence of $\kappa^{(2)}$ on the direction of magnetic fields as
    shown in Fig.~\ref{fig6}(a).
    We vary $\theta$ while maintaining $|\bm{h}|$ uniformly, as shown in Fig.~\ref{fig6}(b).
    Specifically, we set the magnetic fields as $\bm{h}=(0.1\sqrt{2}\cos{\theta}, 0.1\sqrt{2}\sin{\theta}, 0.1)$.
    $\kappa^{(2)}$ acquires significant magnitudes around $\theta=3\pi/4$ and $7\pi/4$,
    corresponding to $\bm{h}=(-0.1, 0.1, 0.1)$ and $\bm{h}=(0.1, -0.1, 0.1)$.
    These $\theta$ values maximize the strength of $D_{1}$ and $D_{3}$.
    In contrast, $\kappa^{(2)}$ is numerically zero
    when $\theta=\pi/4$ and $5\pi/4$ because $D_{1}$ and $D_{3}$ are zero.
    It is noteworthy that the band structures become gapless
    and the Chern number $\nu=0$ when $\theta=0, \pi/2, \pi, 3\pi/2$ and $2\pi$.
    The nonlinear conductivity exhibits anisotropic dependence for specific directions of magnetic fields
    because it vanishes when both $\nu=0$ and $D_{1}=D_{3}=0$.

    \subsection{Influence from flux excitations}
    \begin{figure*}[t!]
        \includegraphics[width=180mm]{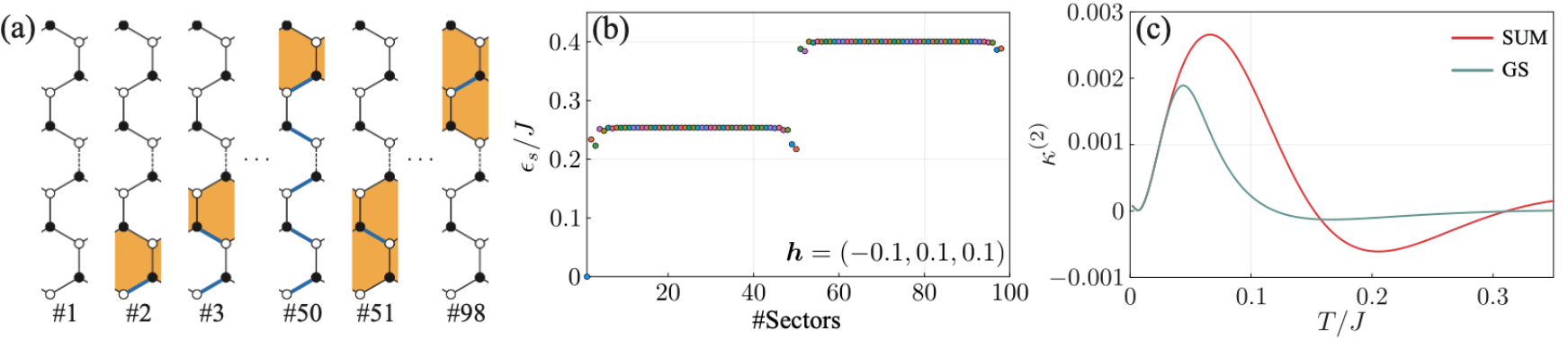}
        \caption{
        (a) Outlines of the sectors sampled in our calculation. The blue bonds represent
        flips of the $Z_{2}$ gauge fields, and the orange plaquettes are excited with localized fluxes.
        Due to a periodic boundary condition in $X$ direction, each of these sectors is a ``striped'' flux sector
        in which fluxes are located in a row along $X$ direction as a whole.
        For a unit cell with $N=100$ sites, 
        \#1 is the zero-flux sector, \#2-\#50 are single-flux sectors and \#51-\#98 are pair-flux (two adjacently excited plaquettes) sectors.
        For single-flux (pair-flux) sectors, the location of the excited plaquette moves to the positive $Y$ direction
        from \#2 to \#50 (from \#51 to \#98).
        (b) The energy difference of single-flux and pair-flux sectors
        which is compared with the zero-flux sector in $\bm{h}=(-0.1, 0.1, 0.1)$.
        (c) Nonlinear conductivity with sampling single-flux and pair-flux sectors
        (SUM: described as Eq.~\eqref{eq:ave})
        and contribution to $\langle \kappa^{(2)} \rangle_{s}$ from zero-flux sector (GS) in $\bm{h}=(-0.1, 0.1, 0.1)$.
        }
        \label{fig7}            
    \end{figure*}

    Finally, we evaluate the influence of flux excitations by sampling excited flux sectors
    using the equation,
    \begin{align}
        \langle \kappa^{(2)} \rangle_{s} = \frac{\sum_{s}\exp(-\epsilon_{s}/T)\kappa^{(2)}_{s}}{\sum_{s}\exp(-\epsilon_{s}/T)}, \label{eq:ave}
    \end{align}
    where $s$ is the index of sectors and $\epsilon_{s}$ is the energy difference
    between the ground-state (zero-flux sector) and each sector $s$.
    Due to our numerical setting, we can only access ``striped'' flux sectors along the $X$ direction.
    We consider single-flux sectors and pair-flux (adjacent) sectors as shown in Fig.~\ref{fig7}(a). 
    Sector \#1 corresponds to the zero-flux sector, sectors \#2-\#50 represent single-flux sectors,
    and sectors \#51-\#98 denote pair-flux sectors.
    For both single-flux and pair-flux sectors, the location of the excited plaquette is shifted
    in the positive $Y$ direction from \#2 to \#50 and from \#51 to \#98, respectively.
    In Fig.~\ref{fig7}(b), we present the energy differences ($\epsilon_{s}$) between each sector and the ground-state.
    The energy level is lower when the excited plaquette locates closer to the edge.
    In Fig.~\ref{fig7}(c), we show the result of the nonlinear conductivity
    with sampling these sectors when $\bm{h}=(-0.1, 0.1, 0.1)$.
    SUM represents the total $\kappa^{(2)}$ considering excited flux sectors weighted
    by the Boltzmann factor as expressed in Eq. \eqref{eq:ave}.
    Although the zero-flux sector is dominant at low temperature,
    other sectors enhance the peak at $T/J\simeq0.07$
    and suppress the peak at $T/J\simeq0.18$ in Fig~\ref{fig5}(b).
    Therefore, the nonlinear conductivity exhibits a certain level of stability
    with respect to flux excitations in low temperatures.

\section{Discussion}\label{sec:Discussion}
    In this section, we quantitatively estimate the magnitudes of the nonreciprocal heat conductivity for realistic systems.
    Considering the case of the $\mathrm{\alpha}$-$\mathrm{RuCl_{3}}$, we adopt the Kitaev interaction
    $J/k_{\mathrm{B}}=80$ K~\cite{Sandilands2015, Nasu2016} and the lattice constants
    $a=6.0$ \AA, $\bar{b}=2.6$ \AA, $c=5.9$ \AA.
    (Here, $\bar{b}$ denotes the average length in the $Y$ direction between sites,
    i.e., the total width of the system $B=N\bar{b}$)~\cite{Johnson2015}.
    For a clean system with $J\tau/\hbar\sim10^{5}$, the conductivities $\tilde{\kappa}^{(1)}$ and $\tilde{\kappa}^{(2)}$ are estimated as
    \begin{align}
        \begin{split}
            \tilde{\kappa}^{(1)}=\frac{a}{\bar{b}c}\frac{k_{\mathrm{B}}J^{2}}{\hbar^{2}}\tau\kappa^{(1)} \sim 10~\mathrm{W/(Km)},\\
            \tilde{\kappa}^{(2)}=\frac{a^{2}}{\bar{b}c}\frac{k_{\mathrm{B}}^{2}J^{2}}{\hbar^{3}}\tau^{2}\kappa^{(2)} \sim 10^{-5}~\mathrm{W/K^{2}}.
        \end{split}
    \end{align}
    The nonlinear conductivity is evaluated to be smaller than $\tilde{\kappa}^{(1)}$
    by approximately the order of relaxation time ($\tilde{\kappa}^{(2)}/\tilde{\kappa}^{(1)}\sim\tau\times10^{2}$ m/K).
    According to $\partial T/ \partial X=160$ K/m~\cite{Kasahara2018, Hirobe2017},
    the ratio of the conductivities is approximated as $\tilde{\kappa}^{(2)}\left(-\frac{\partial T}{\partial X}\right)/\tilde{\kappa}^{(1)}\sim10^{-4}$,
    and each linear and nonlinear component of the thermal current
    (the first and second terms in Eq.~\eqref{eq:current}) is quantitatively estimated as
    $\tilde{\mathcal{J}}^{(1)}\sim10^{4}~\mathrm{W/m^{2}}$ and $\tilde{\mathcal{J}}^{(2)}\sim1~\mathrm{W/m^{2}}$.
    Although
    the nonreciprocal current is orders of magnitude smaller than the conventional contribution,
    this nonreciprocal characteristic would be observable,
    as currently accessible experimental techniques enable us to detect $\kappa_{xx}\sim1~\mathrm{W/Km}$ and $\kappa_{xy}\sim10^{-4}~\mathrm{W/Km}$
    for the Kitaev candidate $\mathrm{\alpha}\text{-}\mathrm{RuCl_{3}}$ at $5~\mathrm{K}$~\cite{Kasahara2018,Yokoi2021}.
    Also, there are several possible suggestion to enhance the nonlinear effect:
    (1) shorten the length of systems in the $Y$ direction,
    (2) increase the steepness of the thermal gradient,
    (3) use a sample with long relaxation time, and so on.
    The nonlinear effects, which are derived from the edge states, may be observed
    by focusing on the $Y$-directional length dependence of the samples.
    Additionally,  measuring the dependence of $\tilde{\mathcal{J}}/(-\frac{\partial T}{\partial X})$ on the thermal gradient would enable us
    to estimate the conductivities on the basis of the intercept and slope.

    \begin{figure*}[t!]
        \includegraphics[width=180mm]{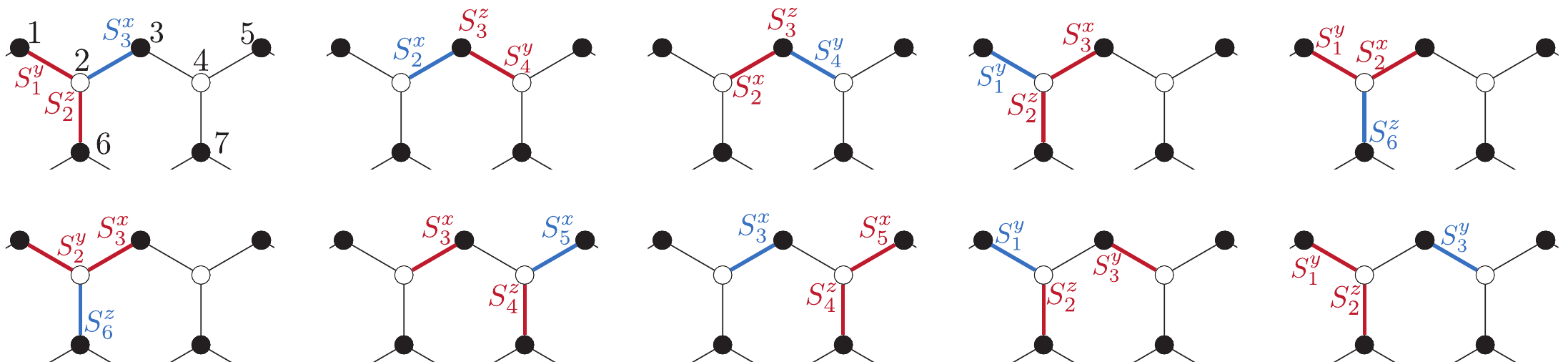}
        \caption{All perturbation processes of the DMI at the upper edge.
        The red and blue bonds represents flips of the $Z_{2}$ gauge fields by the DMI and by Zeeman term.}
        \label{fig8}            
    \end{figure*}
    
    \begin{figure*}[t!]
        \includegraphics[width=180mm]{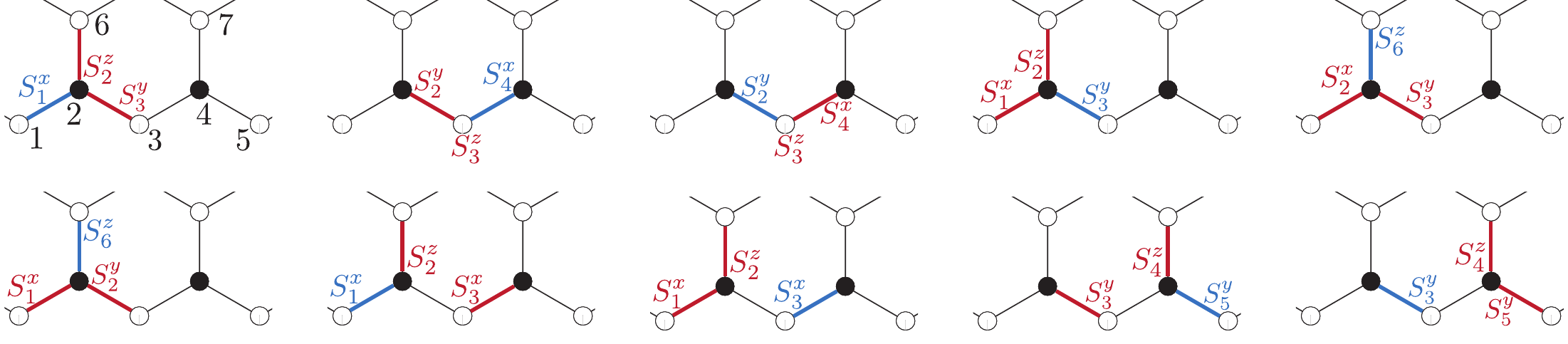}
        \caption{All perturbation processes of the DMI at the lower edge.
        The red and blue bonds represents flips of the $Z_{2}$ gauge fields by the DMI and by Zeeman term.}
        \label{Fig9}            
    \end{figure*}

\section{Conclusion}\label{sec:Conclusion}
    We have theoretically examined nonreciprocal heat transport
    in the Kitaev chiral spin liquid considering open edges.
    The DMI at the edge induces an asymmetry in the Majorana band, thereby leading to nonreciprocal heat transport.
    We stress that the nonreciprocal transport considered here occurs
    even if we take into account the contributions from both upper and lower edges.
    This feature is contrasted to the previous studies which focus on one particular edge~\cite{Yasuda2020}.
    This nonreciprocal heat transport is enhanced in the low temperature region,
    while the dilute flux excitation in the bulk can further enhance its signature.
    Furthermore, there exists the potential to manipulate both of direction and magnitude of
    the nonreciprocal heat current by varying the external magnetic fields,
    corresponding to the sign and the absolute value of $D_{1}=D_{3}=h_{x}D_{x}-h_{y}D_{y}$.
    As a future perspective, it would be interesting to explore how an inhomogeneous potential influences the bulk thermal current~\cite{Pachos2020}.
    Further development of nonreciprocal phenomena is expected in various Majorana systems.

\section*{Acknowledgements}
The authors are grateful to T. Morimoto and Y. Kasahara for invaluable discussion. 
M. O. T. is supported by a Japan Society for the Promotion of Science (JSPS) Fellowship for Young Scientists
and by Program for Leading Graduate Schools: “Interactive Materials Science Cadet Program.”
M. G. Y. is supported by a Japan Science and Technology Agency (JST) PRESTO (Grant No. JPMJPR225B)
and by the Center of Innovation for Sustainable Quantum AI (JST Grant No. JPMJPF2221).
This work is supported by JST CREST Grant No. JPMJCR19T5
and JSPS KAKENHI No. 23K20828, No. JP22H01221, No. JP22J20066 and No. JP22K14005.

\appendix
\section{Perturbation of the DMI}\label{sec:AA}
    The DMI effective Hamiltonian (Eq.~\eqref{eq:effDM}),
    which is obtained by using the perturbation theory for the ground-state of Kitaev spin liquid,
    consists of eight terms,
    \begin{align}
        \begin{split}
            H^{(1)}_{\mathrm{DM}}=H_{\mathrm{D}1}&+H_{\mathrm{D}2}+H_{\mathrm{D}3}+H_{\mathrm{D}4}\\
            &+H_{\mathrm{D}5}+H_{\mathrm{D}6}+H_{\mathrm{D}7}+H_{\mathrm{D}8}.
        \end{split}
    \end{align}
    All processes, which arise from the first order perturbation of the DMI and the magnetic fields,
    at the upper edge are shown in Fig.~\ref{fig8}.
    The red and blue lines represent the bonds flipped by the DMI and the magnetic fields.
    For example, the term $H_{D1}$ comes from four processes as follows
    \begin{align}
        \begin{split}
            H_{\mathrm{D}1}&=H_{\mathrm{D}1(1)}+H_{\mathrm{D}1(2)}+H_{\mathrm{D}1(3)}+H_{\mathrm{D}1(4)}\\
            &=-\frac{1}{4\Delta}(h_{x}D_{x}-h_{y}D_{y})(ic_{1}c_{3}+ic_{2}c_{4}),
        \end{split}
    \end{align}
    where
    \begin{align}
        \begin{split}
            H_{\mathrm{D}1(1)}&=\frac{2!}{\Delta}(-h_{x}S^{x}_{3})(D_{x}S^{z}_{2}S^{y}_{1})\\
            &=-\frac{1}{4\Delta}h_{x}D_{x}ic_{1}c_{3},\\
            H_{\mathrm{D}1(2)}&=\frac{2!}{\Delta}(-h_{y}S^{y}_{1})(-D_{x}S^{z}_{2}S^{x}_{3})\\
            &=\frac{1}{4\Delta}h_{y}D_{y}ic_{1}c_{3},\\
            H_{\mathrm{D}1(3)}&=\frac{2!}{\Delta}(-h_{x}S^{x}_{2})(-D_{x}S^{y}_{4}S^{z}_{3})\\
            &=-\frac{1}{4\Delta}h_{x}D_{x}ic_{2}c_{4},\\
            H_{\mathrm{D}1(4)}&=\frac{2!}{\Delta}(-h_{y}S^{y}_{4})(D_{y}S^{x}_{2}S^{z}_{3})\\
            &=\frac{1}{4\Delta}h_{y}D_{y}ic_{2}c_{4},
        \end{split}
    \end{align}
    the site index is same as Fig.~\ref{fig8}.
    The ``$c$ to $b^{z}$ hopping" term $H_{D5}$ is obtained in the same way
    \begin{align}
        \begin{split}
            H_{\mathrm{D}5}=-\frac{1}{4\Delta}2h_{x}D_{y}ib^{z}_{3}c_{5},
        \end{split}
    \end{align}
    where
    \begin{align}
        \begin{split}
            H_{\mathrm{D}5(1)}&=\frac{2!}{\Delta}(-h_{x}S^{x}_{5})(-D_{y}S^{z}_{4}S^{x}_{3})\\
            &=-\frac{1}{4\Delta}h_{x}D_{y}ib^{z}_{3}c_{5},\\
            H_{\mathrm{D}5(2)}&=\frac{2!}{\Delta}(-h_{x}S^{x}_{3})(-D_{y}S^{z}_{4}S^{x}_{5})\\
            &=-\frac{1}{4\Delta}h_{x}D_{y}ib^{z}_{3}c_{5}.\\
        \end{split}
    \end{align}
    For lower edges, the terms $H_{\mathrm{D}3}, H_{\mathrm{D}4}, H_{\mathrm{D}7}$
    and $H_{\mathrm{D}8}$ are obtained as well as the upper edge, following Fig.~\ref{Fig9}.
    In this way, the DMI effective Hamiltonian is expressed by Eqs.~\eqref{eq:effDM}~and~\eqref{eq:DMhop}.

    \begin{figure}[t!]
        \includegraphics[width=\columnwidth]{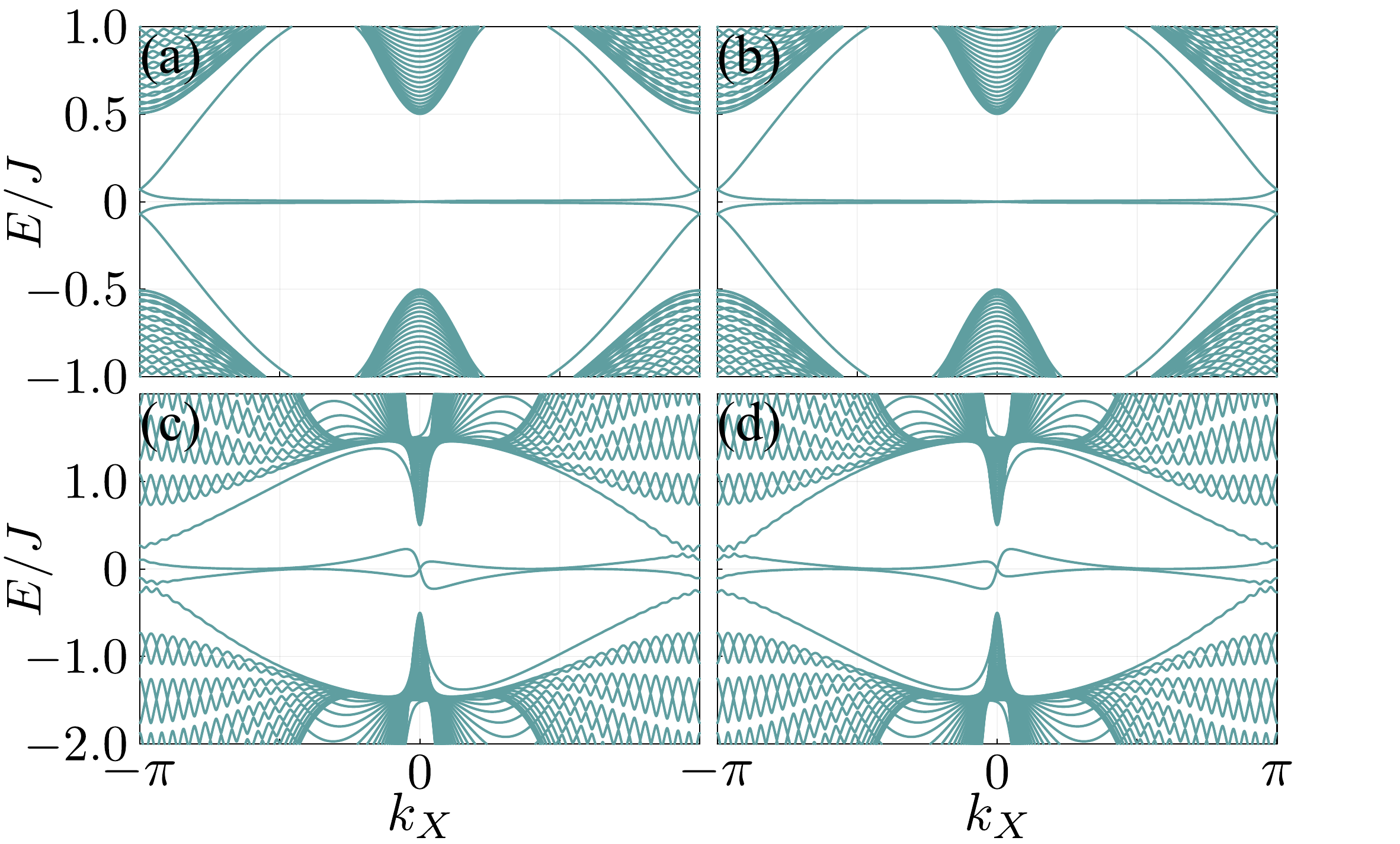}
        \caption{The band structures when (a) $\bm{h}=(0.1, 0.1, 0.1)$, $\bm{D}=(0,0,0)$; (b) $\bm{h}=(0.1, -0.1, 0.1)$, $\bm{D}=(0,0,0)$;
                (c) $\bm{h}=(0.2, -0.2, 0.2)$, $\bm{D}=(0.1,0.1,0.1)$; and (d)$\bm{h}=(-0.2, 0.2, 0.2)$, $\bm{D}=(0.1,0.1,0.1)$.
                (a) and (b), cases without the DMI, are numerically identical and symmetric. (c) and (d) are asymmetric and inverted.
                }
        \label{fig10}
    \end{figure}

    \begin{figure}[b!]
        \includegraphics[width=\columnwidth]{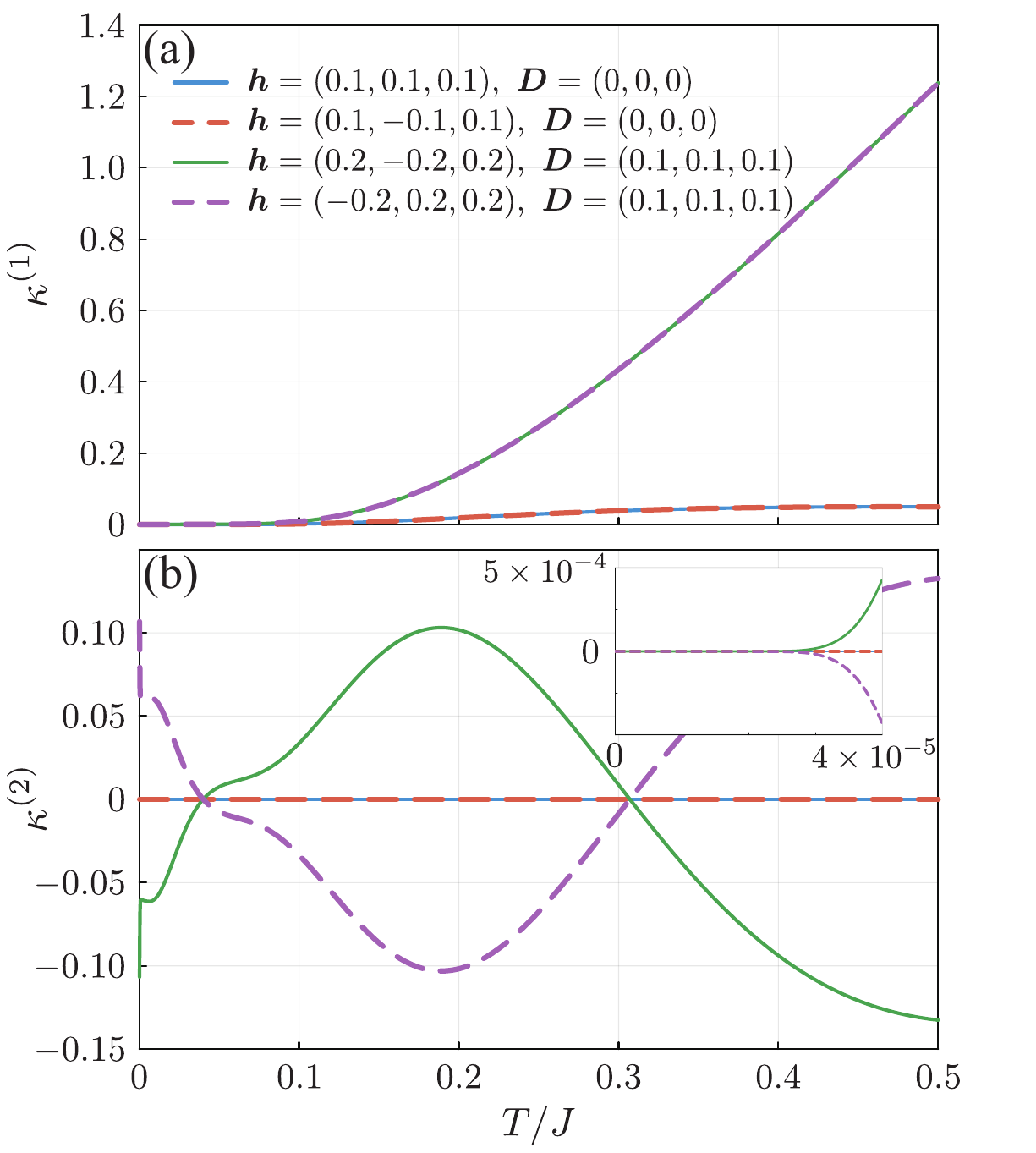}
        \caption{In a variety of parameters, (a) the linear conductivity $\kappa^{(1)}$,
                (b) the nonlinear conductivity $\kappa^{(2)}$.
                Each of blue, red, green and purple lines represents $\bm{h}=(0.1, 0.1, 0.1)$, $\bm{D}=(0,0,0)$; $\bm{h}=(0.1, -0.1, 0.1)$, $\bm{D}=(0,0,0)$;
                $\bm{h}=(0.2, -0.2, 0.2)$, $\bm{D}=(0.1,0.1,0.1)$; and $\bm{h}=(-0.2, 0.2, 0.2)$, $\bm{D}=(0.1,0.1,0.1)$.
                The inset shows the nonlinear conductivity at the extreme low temperature.
                }
        \label{fig11}
    \end{figure}

\section{Band structures and heat conductivities on various parameters}\label{sec:AB}

In this section, we present band structures and conductivities
for various sets of a magnetic field $\bm{h}$ and the DMI $\bm{D}$,
particularly for the cases where $\bm{h}=(0.1, 0.1, 0.1)$, $\bm{D}=(0,0,0)$; $\bm{h}=(0.1, -0.1, 0.1)$, $\bm{D}=(0,0,0)$;
$\bm{h}=(0.2, -0.2, 0.2)$, $\bm{D}=(0.1,0.1,0.1)$; and $\bm{h}=(-0.2, 0.2, 0.2)$, $\bm{D}=(0.1,0.1,0.1)$.

In Figs.~\ref{fig10}(a)~and~\ref{fig10}(b), we show the band structures
with different magnetic fields but without the DMI,
which are numerically identical and symmetric.
In Figs.~\ref{fig10}(c)~and~\ref{fig10}(d), we show the effects of stronger magnetic fields, compared with Figs.~\ref{fig3}(a)~and~\ref{fig3}(b), on band structures.
The increased magnetic field intensifies the asymmetry and widens the gap at $k_{X}=\pm\pi$.
Notably, the transformation ``$h_{x} \rightarrow -h_{x}$ and $h_{y}\rightarrow -h_{y}$" inverts the band structures, here again.

In Figs.~\ref{fig11}(a)~and~\ref{fig11}(b), we show the linear and nonlinear conductivities.
For the blue and red line, i.e., in the case of $\bm{D}=0$, the liner conductivities retain as same magnitude as those in Fig.~\ref{fig5}(a),
and the nonlinear conductivities exhibit no nonzero value.
In contrast, for the green and purple lines, both the linear and nonlinear conductivities are enhanced by factors on the order of $10$ to $10^{2}$,
compared with those in Figs.~\ref{fig5}(a)~and~\ref{fig5}(b).
Furthermore, for the large magnetic field, the temperature dependence of the heat conductivities at low temperatures are qualitatively different
from the case with smaller magnetic field.

\bibliography{paper}

\end{document}